# Kernel-smoothed proper orthogonal decomposition (KSPOD)-based emulation for prediction of spatiotemporally evolving flow dynamics

Yu-Hung Chang [a,*], Liwei Zhang [a,**], Xingjian Wang [a,†], Shiang-Ting Yeh [a,††], Simon Mak [b,‡], Chih-Li Sung [b,‡], C. F. Jeff Wu [b,§], Vigor Yang [a,¶]

[a] *School of Aerospace Engineering, Georgia Institute of Technology, Atlanta, Georgia, USA*

[b] *School of Industrial and Systems Engineering, Georgia Institute of Technology, Atlanta, Georgia, USA*[*]

## Abstract

This interdisciplinary study, which combines machine learning, statistical methodologies, high-fidelity simulations, and flow physics, demonstrates a new process for building an efficient surrogate model for predicting spatiotemporally evolving flow dynamics. In our previous work, a common-grid proper-orthogonal-decomposition (CPOD) technique was developed to establish a physics-based surrogate (emulation) model for prediction of mean flowfields and design exploration over a wide parameter space. The CPOD technique is substantially improved upon here using a kernel-smoothed POD (KSPOD) technique, which leverages kriging-based weighted functions from the design matrix. The resultant emulation model is then trained using a dataset obtained through high-fidelity simulations.

As an example, the flow evolution in a swirl injector is considered for a wide range of design parameters and operating conditions. The KSPOD-based emulation model performs well, and can faithfully capture the spatiotemporal flow dynamics. The model enables effective design surveys utilizing high-fidelity simulation data, achieving a turnaround time for evaluating new design points that is 42,000 times faster than the original simulation.

*Co-first author, Graduate Student, School of Aerospace Engineering
**Co-first author, Research Engineer, School of Aerospace Engineering
†Research Engineer, School of Aerospace Engineering
†† Graduate Student, School of Aerospace Engineering
‡Graduate Student, School of Industrial and Systems Engineering
§Professor and Coca-Cola Chair in Engineering Statistics, School of Industrial and Systems Engineering
¶Corresponding author, William R. T. Oakes Professor and Chair, School of Aerospace Engineering,
vigor.yang@aerospace.gatech.edu

# I. Introduction

The purpose of this work is to develop an accurate and efficient surrogate model for prediction of spatiotemporally evolving flow dynamics. The study involves computational fluid dynamics (CFD), reduced-basis modeling, statistics, and machine learning. As a demonstration case, the flow evolution in a swirl injector is presented.

For design assessment, physical experiments can be extremely expensive and time-consuming, especially for complex systems operating over a wide range of conditions. Moreover, it is hard to gain insight into underlying physicochemical mechanisms through measurements using currently available experimental techniques. To better capture flow characteristics and identify design attributes, one solution is high-fidelity modeling and simulations such as large-eddy simulation (LES). The LES framework employed in the present work is capable of dealing with fluid flow and combustion dynamics over the entire range of thermodynamic states [1-5]. These simulations, however, are computationally expensive and impractical for use as a primary tool to survey the design space; an axisymmetric simulation of flow evolution in a simplex swirl injection with LES-grade resolution, for instance, may take about 100,000 CPU hours on the hexa-core AMD Opteron Processor 8431. The traditional trial-and-error based design practice is no longer practical. To enable the use of high-fidelity simulations for design evaluation, an effective model must be incorporated into the design process [6,7].

The first step toward the development of an emulation (surrogate) model is Design of Experiments (DoE), which achieves more realistic computation timelines for building a database. Here we consider a swirl injector as a demonstration example [8], as shown schematically in Fig. 1. The detailed flow characteristics have been previously explored using LES techniques [9, 10]. DoE can be formulated based on several key geometric parameters and their respective ranges of consideration. The present study focuses on the effects of these geometric parameters (that is, location and width of the tangential entry and injection angle) on the injector performance, as measured by the thickness and spreading angle of the liquid film at the exit of the injector [7-10]. The total sample size is determined using a 10*d* rule-of-thumb described by Loeppky et al. [11], which recommends ten simulations per design parameter where *d* means the total number of design parameters. This approach substantially reduces the number of total sample points required to survey the design space.

The second step is the creation of a database with sufficient information to allow for a survey of the design space. This can be achieved by performing an LES-based high-fidelity simulation at the selected design points. For spatiotemporally evolving flows, however, the resulting database is too large to be handled effectively. Identification of dominant flow structures and reduction of the "big data" becomes essential for building the emulation model [7].

The final step requires the combination of POD and kriging methodologies. Kriging is a powerful machine-learning tool for interpolation and prediction [12]. The concept of kriging is to model unobserved responses using a Gaussian Process (GP) governed by a preset covariance function. The response surface of the training model can be evaluated via data-tuned weights to radial basis functions centered at observed points. In the present study, each simulation contains over 40,000 numerical grid points in the spatial domain. Kriging is required for each point if information for each point is to be modeled. To reduce the data size, proper orthogonal decomposition (POD) [13] has been incorporated into the prediction model (emulator).

In our previous work [6, 7], common-grid POD (CPOD) was implemented to build an emulation model. The method successfully predicts mean flow structures for swirl injectors with a broad range of geometric dimensions. To improve the prediction accuracy of flow evolution in the entire spatial domain and associated flow dynamics, Kernel-smoothed POD (KSPOD) is developed here. Two main assumptions must be addressed. First, the physics extracted by POD modes in different cases are similar under the same rank, as determined by the POD energy. Second, the dominant modes capturing similar physics are transferred with similar phase and those physics are retained through the kriging-weighted averaging based on the new design. KSPOD achieves three goals: (i) capturing turbulent flow dynamics; (ii) accurately predicting results as verified through quantitative comparisons with simulation results; (iii) yielding predictions with short turnaround times. The KSPOD-based emulator requires only about 0.02 CPU hour on an Intel Xeon Processor E5-1650 V4 to predict the flowfield at a new design point for the problem presented in the present study. The overall computation time, including data loading and training, is about 25 CPU hours (on an Intel Xeon Processor E5-1650 V4) to predict the flow evolution at a new design point for a time duration of 10 ms with 1000 snapshots. For comparison, the corresponding LES calculation takes about 100,000 CPU hours (on the hexa-core AMD Opteron Processor 8431) for a spatial domain of over 40,000 numerical grid points.

The work described in this paper produces a high-fidelity surrogate modeling technique for efficient prediction of complex flowfields over a broad range of operating conditions and geometric parameters. The paper is structured as follows. Section II introduces the swirl injector configuration, sample points designated by the DoE, the numerical framework, the emulation model, and the overall design process. Section III discusses the simulation and emulation results. The emulator is systematically assessed based on performance metrics, relative error, and spectral information on simulated and predicted flowfields. Section IV concludes with directions for future work.

# II. Methodology

## Swirl injector configuration

Figure 1 shows a schematic of the swirl injector of concern. Liquid oxygen (LOX) is tangentially introduced into the injector, and develops a swirling film that is attached to the wall due to centrifugal force [8, 14]. Conservation of angular momentum results in a hollow gaseous core in the center region. The liquid film exits the injector as a thin conical sheet and subsequently undergoes atomization into droplets. The flow dynamics in this type of device under supercritical conditions have been extensively studied using LES techniques [9, 10].

The selection of design variables is dependent upon system requirements. In our previous study [6,7], the design consisted of five parameters: injector length, $L$, injector radius, $R_n$, tangential inlet width, $\delta$, tangential injection angle, $\theta$, and distance between the inlet and headend, $\Delta L$. The sample points were chosen based on the Maximum Projection (MaxPro) [15] technique across a broad design space. The injector length covers a range of 22-93 mm. Sensitivity analysis is used to identify the most significant parameters dictating injector performance. Our previous work [7] shows that the injection width ($\delta$) is the most important parameter in the determination of the spreading angle of the liquid film. The tangential inlet angle ($\theta$) and the injection width ($\delta$) significantly affect the liquid film thickness, while the injector length $L$ and radius $R_n$ play minor roles. Thus, the present work focuses on the injection width $\delta$ , and angle $\theta$, and the distance between the inlet and headend $\Delta L$.

Table 1 tabulates the baseline geometry and operating conditions, including the LOX inlet temperature $T_{in}$, ambient temperature $T_\infty$, ambient pressure $p_\infty$, and mass flow rate $\dot{m}$. Another parameter is the geometric constant $K$, a nondimensional parameter that can be used to evaluate the flow characteristics of swirl injectors, namely the liquid film thickness and spreading angle. The former controls film atomization and the latter mixing efficiency [8]. The geometric constant takes the following definition.

$$K = A_n R_{in} / A_{in} R_n \tag{1}$$

where $A_n$ denotes the cross-sectional area of the injection exit and $A_{in}$ the total inlet area. The geometric constant is an indicator of the swirl strength. When the value is high, a large angular momentum is present in the liquid film, leading to a wide spreading angle.

Table 2 shows the design space and the range for each design variable. The distance between the inlet and headend, $\Delta L$, is decided by a rule of thumb to be 1.5-2 times the injector width [8]. This is an optimal location determined from a tradeoff study to avoid (1) excessive viscous losses when the injection slit is too close to the headend and (2) low frequency oscillations due to the presence of a large recirculation zone if the inlet is too far from the headend. The design spaces of injection width $\delta$ and angle $\theta$ are decided by the desired range of spreading angle (50-62°) and film thickness (0.66-1.50 mm). With these numbers and the geometric constant, the ranges of $\delta$ and $\theta$ can be estimated.

Table 1. Baseline geometry and operating conditions

| $R$ (mm) | $R_{in}$ (mm) | $L$ (mm) | $\dot{m}$ (kg/s) | $T_{in}$ (K) | $T_\infty$ (K) | $p_\infty$ (MPa) |
|---|---|---|---|---|---|---|
| 4.50 | 0.85 | 25 | 0.17 | 120 | 300 | 10 |

Table 2. Design space

| Design Variable | $\theta$ (deg) | $\delta$ (mm) | $\Delta L$ (mm) |
|---|---|---|---|
| Design Range | 35.0-62.2 | 0.27-1.53 | 0.85-3.40 |

## High-fidelity simulation

The theoretical formulation for high-fidelity simulations, which treats supercritical fluid flows and combustion over the entire range of fluid thermodynamic states of concern, is described in detail in [2,3]. Turbulence closure is achieved by means of LES techniques. The effects of subgrid-scale motion are represented by an improved Smagorinsky eddy-viscosity model. Thermodynamic properties are evaluated according to fundamental thermodynamics theories and a modified Soave-Redlich-Kwong (SRK) equation of state (EOS). The Takahashi method calibrated for high-pressure conditions is employed to obtain the mass diffusivity. Transport properties are evaluated using extended corresponding-state principles.

The numerical framework is based on a preconditioning scheme with a unified treatment of general-fluid thermodynamics [16,17]. It applies a density-based, finite-volume methodology, along with a dual-time-step integration technique [18]. A second-order backward difference is used to accomplish temporal discretization; and a four-step Runge-Kutta scheme is applied to integrate the inner-loop pseudo-time term. A fourth-order central difference scheme in generalized coordinates is used to obtain spatial discretization. Fourth-order matrix dissipation is taken to assure numerical stability and minimum contamination of the solution. Lastly, a multi-block domain decomposition technique associated with the message passing interface technique for parallel computing is applied to optimize computation speed.

## Design of Experiments (DoE)

The DoE methodology is a statistical approach for careful selection of input variables for a given design space. DoE facilitates the design process and reduces the number of total sample points required to efficiently explore the design space. With prior knowledge of the major contributing geometric parameters, the sample size is determined based on the $10d$ rule-of-thumb [11], where $d$ denotes the total number of design parameters. The present study considers three design variables ($\theta$, $\delta$, and $\Delta L$). Although MaxPro provides good space-filling properties on both the design space and its projections [15], MaxPro does not provide a sequential design capability, giving optimal space-filling performance in batches. To that end, Sliced Latin Hypercube Design (SLHDs) [19-21] is selected. In SLHDs, the space-filling performance of the design points in each slice is optimal. The overall design matrix contains five slices, and each slice includes six design points. Figure 2 shows the two-dimensional projections of the design points categorized by different slices. Each case takes about 100,000 CPU hours for high-fidelity simulation to obtain statistically significant data. A total of 1000 snapshots spanning 10 ms are acquired after the flowfield reaches its stationary state (~12 ms). The snapshots are subsampled every 20 temporal iterations, each with a time step of $0.5\mu s$. A temporal resolution of 50 kHz is achieved, according to the Nyquist criterion.

## Kernel-Smoothed POD

This section introduces kernel-smoothed proper orthogonal decomposition (KSPOD), which combines statistical modeling with a data-reduction method to obtain a reduced-basis model. The proposed KSPOD method can be viewed as a generalization of the POD used for flow emulation. POD decomposes the flowfield into an expansion consisting of spatial eigenfunctions, known as POD modes, and corresponding time-varying coefficients. Such a decomposition can be written in the following form.

$$f(\boldsymbol{u}, t) \approx \sum_{k=1}^{K} \beta^k(t)\phi^k(\boldsymbol{u}) \tag{2}$$

where $f(\boldsymbol{u}, t)$ is the simulated flowfield at spatial location $\boldsymbol{u}$ and time $t$, and $\beta^k(t)$ and $\phi^k(\boldsymbol{u})$ represent the time-varying coefficient and basis function for the $i$-th mode, respectively. As indicated in Eq. (2), the expansion is truncated with the first $K$ terms, where $K$ is chosen such that the reconstructed flowfield retains a desired degree of accuracy. In practice, the time-varying coefficients and basis functions are obtained through an eigen-decomposition of the inner-product of a flowfield variable [13]. Equation (2) can be viewed as the optimal decomposition of $f(\boldsymbol{u}, t)$ using a basis expansion of $K$ terms.

From a physics perspective, POD provides valuable insight into the important physics present in the flowfield. The basis function, or mode shape, $\phi^k(\boldsymbol{u})$ can be interpreted as spatial distribution of the fluctuation field of a given flow variable (such as pressure, density, temperature, and velocity components). It represents the dominant coherent structure, such as acoustic waves in the system [22]. A spectral analysis of the POD coefficients can be performed to identify flow periodicity and characteristic frequencies for hydrodynamic and acoustic instabilities. The index for the basis-function expansion in Eq. (2) is determined by the rank of the energy content in the eigen-decomposition calculation, and suggests which flow structure is more prevalent. The first few terms in the expansion represent more energy-containing structures, and the remaining terms represent increasingly weaker flow features.

The goal here is to employ the flow features extracted using POD within a statistical framework, allowing the training of an emulator for flow prediction. To this end, we employ a machine-learning technique called Gaussian process regression, also known as kriging, to predict POD modes and time-varying coefficients at a new design setting. We begin with a brief introduction into the mathematical formulation behind kriging, and then describe how such a model is incorporated into the KSPOD framework.

Kriging is a powerful learning technique that leverages a Gaussian process regression model to learn the structure of an unknown function by sampling this function at specific points. The mathematical framework for kriging is established as described in [23, 24]. We assume the unknown function of interest, $Y(\boldsymbol{x}), \boldsymbol{x} \in \mathbb{R}^d$, is a realization from the stochastic process

$$Y(\boldsymbol{x}) = \mu + Z(\boldsymbol{x}) \tag{3}$$

where $\boldsymbol{x}$ is a $d$-dimensional vector, $\mu$ is the mean of the process, and $Z(\boldsymbol{x})$ is a zero-mean Gaussian process with $Var\{Z(\boldsymbol{x})\} = \sigma^2$ and correlation function

$$Corr\left[Z(\boldsymbol{x}_i), Z(\boldsymbol{x}_j)\right] = R\left(\boldsymbol{x}_i, \boldsymbol{x}_j\right) \tag{4}$$

where $R(\boldsymbol{x}_i, \boldsymbol{x}_j)$ denotes the correlation between the random variables $Z(\boldsymbol{x}_i)$ and $Z(\boldsymbol{x}_j)$. Following common practice [25], we employ the squared-exponential correlation function:

$$R(\boldsymbol{x}_i, \boldsymbol{x}_j) = exp\left[-\sum_{k=1}^{d} \theta_k (x_{ik} - x_{jk})^2\right] \quad (5)$$

where $x_{ik}$ is the $k$-th element of $\boldsymbol{x}_i$.

A key advantage of kriging is that a closed-form expression can be obtained for predicting the unknown function $Y(\boldsymbol{x})$ at unobserved locations. Suppose the function of interest $Y$ is observed at the design setting $\{\boldsymbol{x}_i\}_{i=1}^{n}$, the observation vector then becomes $\boldsymbol{y} = [Y(\boldsymbol{x}_1), \ldots, Y(\boldsymbol{x}_n)]^T$. Having observed $\boldsymbol{y}$, the conditional mean of the process at a new point $\boldsymbol{x_{new}}$ is given by:

$$\hat{Y}(\boldsymbol{x_{new}}) = \mathbb{E}[Y(\boldsymbol{x}_{new})|\boldsymbol{y}] = \hat{\mu} + \boldsymbol{r}^T \boldsymbol{R}^{-1}(\boldsymbol{y} - \boldsymbol{1}_n \hat{\mu}) \quad (6)$$

where $\hat{\mu} = \boldsymbol{1}_n{}^{\boldsymbol{T}} \boldsymbol{R}^{-1} Y / \boldsymbol{1}_n{}^{\boldsymbol{T}} \boldsymbol{R}^{-1} \boldsymbol{1}_n$ is the estimated value of $\mu$, and $\boldsymbol{1}_n$ is an $n$-vector of 1's. Here, $\boldsymbol{R}$ is an $n \times n$ matrix whose $(i, j)$-th entry is $R(\boldsymbol{x_i}, \boldsymbol{x_j})$, and $\boldsymbol{r}$ is an $n$-vector whose $i_{th}$ entry is $R(\boldsymbol{x_{new}}, \boldsymbol{x_i})$. A more detailed derivation of Eq. (6) can be found in [25, 26].

While the predictor in Eq. (6) is easy to evaluate when the desired function $Y$ is a scalar function, it becomes much more difficult to evaluate for the problem at hand, where the desired function involves spatiotemporal evolution. For example, in the present study of swirling flow dynamics, there are over 400,000 grid points and 1,000 time-steps for each simulation case. Performing kriging for each grid point and time step would be impractical and time consuming. From a statistical perspective, the use of separate kriging models over each grid point and time step also leads to a serious problem of over-parametrization (as each model requires $d$ correlation parameters), which then results in poor prediction performance of the trained model. In light of these challenges, and inspired by the fact that the numerical grid system remains the same for all simulated cases, we introduce an improved kriging-based model that combines the POD information from each case in the form of a "weighting number."

The key idea in KSPOD is to apply the kriging equation, Eq. (6), to predict the weight of each POD mode at a new design setting. To this end, the observations $\boldsymbol{y}$ are now taken to be the unit vector $\boldsymbol{e_i}$, where $\boldsymbol{e_i}$ is an $n$-vector with 1 in its $i$-th element and 0 elsewhere. Intuitively, this quantifies the fact that the POD information extracted in the $i$-th design setting corresponds to only that setting and not the other $n$-1 settings. With this in mind, the resulting predictor in Eq. (6) can then be viewed as the predicted weight for that particular POD term at a new design setting $\boldsymbol{x}_{new}$, denoted as $\widehat{w}_{new,i}$ This procedure is repeated for each of the $n$ unit vectors $(\boldsymbol{e_i})_{i=1}^{n}$, from which the $n$ weighting numbers $\left(\widehat{w}_{new,i}\right)_{i=1}^{n}$ can be obtained. They are subsequently used to predict the new POD modes and coefficients through a weighted average of the extracted modes and coefficients at the new design settings.

The algorithm in Table 3 outlines the detailed steps in the KSPOD algorithm. First, POD is performed for each simulated case to extract coherent structures. Next, the coefficients of POD modes are trained by ordinary kriging models using the Gaussian kernel in Eq. (5) as the correlation function, with the correlation parameter $\theta$ tuned using maximum-likelihood estimation, as implemented in the R package 'GPfit' [27]. The predictive function can be constructed based on Eq. (6). The weighting numbers are also trained using the procedure described above. Finally, the POD modes and coefficients are predicted, and are used to "reconstruct" the flowfield at the new design setting $\boldsymbol{x}_{new}$.

As in any physical or statistical model, there are implicit assumptions. First, by predicting the $i$-th POD term of the new design setting using only the information for the $i$-th POD terms extracted from observed design settings, we assume the ranking of the extracted flow physics from POD (corresponding to the rank of its corresponding expansion term in Eq. (2)) to be invariant over different geometry settings.

In other words, the flow feature for the first POD mode corresponds to the same coherent structure over all $n$ simulated design settings. The same holds true for the subsequent modes. Second, for the methodology to work, design settings whose dominant POD modes capture similar physics and dynamics should be clustered together for training. As information can be damped or even diminished during the training process, clustering the data and using cases with similar phase content can mitigate the problem.

Table 3. Algorithm

| ALGORITHM: Kernel-smoothed POD | | |
|---|---|---|
| DATA: | | For each design setting in $\{x_i\}_{i=1}^{n}$, provide the flow evaluation at each spatial location and time-step $f(x_i, u_j, t_q)$, where $\{u_j\}_{j=1}^{J}$ is the spatial location and $\{t_q\}_{q=1}^{m}$ is the time-step. |
| TRAINING: | *Step 1:* | For each design setting $x_i$, perform a POD and write it as $f(x_i, u_j, t_q) = \sum_{k=1}^{K} \beta^k(x_i, t_q)\phi^k(x_i, u_j)$. |
| | *Step 2:* | For each time-step $t_q$ and each mode $k$, perform an ordinary kriging procedure on $\{\beta^k(x_1, t_q), \dots, \beta^k(x_n, t_q)\}$ with inputs $\{x_1, \dots, x_n\}$. The predictive function at an untried setting $x_{new}$ is $\hat{\beta}^k(x_{new}, t_q)$. |
| | *Step 3:* | For $i = 1, \dots, n,$ perform an ordinary kriging procedure on $e_i$ with inputs $\{x_1, \dots, x_n\}$. The predictive function at an untried setting $x_{new}$ is $\widehat{w}_i(x_{new})$. Therefore, for each spatial location $u_j$ and each mode $k$, the predictive function of $\phi^k(x_{new}, u_j)$ is $\hat{\phi}^k(x_{new}, u_j) = \sum_{i=1}^{n} \widehat{w}_i(x_{new})\phi^k(x_i, u_j) / \sum_{i=1}^{n} \widehat{w}_i(x_{new})$ . |
| PREDICTION: | | At an untried setting $x_{new}$, compute $\hat{\beta}^k(x_{new}, t_q), q = 1, \dots, m; k = 1, \dots, K$, and $\hat{\phi}^k(x_{new}, u_j), j = 1, \dots, J; k = 1, \dots, K$, Then$f(x_{new}, u_j, t_q) = \sum_{k=1}^{K} \hat{\beta}^k(x_{new}, t_q)\hat{\phi}^k(x_{new}, u_j)$, where $j = 1, \dots, J$ and $q = 1, \dots, m$. |

The method is termed kernel-smoothed POD, because the kriging here does not apply the weighting number $\widehat{w}_{new,i}$ directly to the flowfield. The flow structures within the flowfield are a combination of waves with different frequencies, amplitudes, and phases. If the weighting number is used on the flowfield directly, the two datasets may cancel each other during the regression process, thereby losing useful information. The phase difference can be observed in POD modes as well. Application of weighting functions to POD modes with a kernel-smoothed algorithm can avoid the phase cancellation and retain important flow physics.

Once the emulator model is trained, it can be used with the KSPOD model to predict the flow evolution at a new design point. The computational cost is reduced by several orders of magnitude. For the demonstration case in the present study, high-fidelity simulations take around 100,000 CPU hours for each case. The trained model can evaluate a new flowfield in 0.02 CPU hours.

# III. Results and Discussion

## High-fidelity simulation results

The LES-based high-fidelity numerical framework described earlier was implemented for the 30 cases selected by SLHDs for the three design parameters, which are decided by the sensitivity analysis using a first-order Monte Carlo estimation of Sobol' indices [7]. The sensitivity analysis was performed based on the desired liquid-film thickness and spreading angle. The three chosen parameters are significantly influenced by the inlet velocity, $u_{in}$, which ranges from 5.71 to 40.43 m/s as listed in Table 4. The 30 training cases are roughly classified into four groups in terms of $u_{in}$ (m/s) as follows: Cluster A with $u_{in} < 10$; Cluster B with $10 \leq u_{in} < 18$; Cluster C with $18 \leq u_{in} < 25$; and Cluster D with $u_{in} > 25$.

Overall, there are 15 cases in Cluster A, 8 in Cluster B, 5 in Cluster C, and 2 in Cluster D. Figure 3 shows two sample snapshots of the density field for each cluster. Variations of film thickness and spreading angle are observed. Cluster A has the slowest inlet velocity and generates thicker film and smaller spreading angle. The film thickness decreases and the spreading angle increases between Clusters

B and C; the smallest film thickness and largest spreading angle appear in Cluster D, where there is the highest inlet velocity.

Table 4. Design matrix and associated inlet velocity information

| Case | $\delta$ (mm) | $\theta$ (deg) | $\Delta L$ (mm) | $u_{in}$ (m/s) | $u_r$ (m/s) | $u_\theta$ (m/s) | $K$ | *Cluster* |
|---|---|---|---|---|---|---|---|---|
| 1 | 0.28 | 57.92 | 1.59 | 40.43 | 21.47 | 34.26 | 7.44 | D |
| 2 | 0.63 | 40.81 | 1.93 | 12.35 | 9.35 | 8.07 | 1.64 | B |
| 3 | 0.82 | 52.39 | 0.96 | 11.79 | 7.20 | 9.34 | 1.98 | B |
| 4 | 1.10 | 32.76 | 2.57 | 6.42 | 5.40 | 3.47 | 0.69 | A |
| 5 | 1.12 | 51.88 | 3.21 | 8.58 | 5.30 | 6.75 | 1.43 | A |
| 6 | 1.52 | 46.85 | 2.23 | 5.71 | 3.90 | 4.16 | 0.86 | A |
| 7 | 0.38 | 37.29 | 1.64 | 19.53 | 15.54 | 11.83 | 2.37 | C |
| 8 | 0.51 | 52.89 | 2.15 | 19.35 | 11.67 | 15.43 | 3.27 | C |
| 9 | 0.78 | 43.33 | 3.12 | 10.43 | 7.58 | 7.15 | 1.46 | B |
| 10 | 1.03 | 33.76 | 0.87 | 6.89 | 5.73 | 3.83 | 0.76 | A |
| 11 | 1.26 | 49.37 | 1.72 | 7.19 | 4.68 | 5.46 | 1.14 | A |
| 12 | 1.39 | 60.44 | 2.61 | 8.63 | 4.26 | 7.51 | 1.65 | A |
| 13 | 0.47 | 54.40 | 2.74 | 21.87 | 12.73 | 17.78 | 3.80 | C |
| 14 | 0.68 | 38.80 | 2.53 | 11.25 | 8.77 | 7.05 | 1.42 | B |
| 15 | 0.74 | 48.36 | 1.89 | 12.06 | 8.02 | 9.02 | 1.88 | B |
| 16 | 0.93 | 33.26 | 1.47 | 7.63 | 6.38 | 4.18 | 0.83 | A |
| 17 | 1.22 | 42.82 | 0.91 | 6.60 | 4.84 | 4.49 | 0.92 | A |
| 18 | 1.35 | 57.42 | 3.17 | 8.15 | 4.39 | 6.87 | 1.49 | A |
| 19 | 0.32 | 58.43 | 2.27 | 35.58 | 18.63 | 30.31 | 6.60 | D |
| 20 | 0.59 | 34.77 | 1.13 | 12.19 | 10.01 | 6.95 | 1.38 | B |
| 21 | 0.84 | 49.87 | 2.83 | 10.89 | 7.02 | 8.32 | 1.74 | B |
| 22 | 0.99 | 44.33 | 1.76 | 8.35 | 5.97 | 5.84 | 1.20 | A |
| 23 | 1.20 | 37.79 | 3.08 | 6.24 | 4.93 | 3.82 | 0.77 | A |
| 24 | 1.45 | 55.41 | 1.55 | 7.17 | 4.07 | 5.90 | 1.27 | A |
| 25 | 0.40 | 36.28 | 2.32 | 18.27 | 14.73 | 10.81 | 2.16 | C |
| 26 | 0.49 | 51.38 | 1.42 | 19.51 | 12.18 | 15.24 | 3.21 | C |
| 27 | 0.72 | 53.39 | 3.29 | 13.84 | 8.25 | 11.11 | 2.36 | B |
| 28 | 0.95 | 40.31 | 1.17 | 8.18 | 6.24 | 5.29 | 1.07 | A |
| 29 | 1.24 | 59.43 | 1.98 | 9.36 | 4.76 | 8.06 | 1.76 | A |
| 30 | 1.37 | 43.83 | 2.78 | 5.99 | 4.32 | 4.15 | 0.85 | A |

Note: Cases 1-6 are on Slice 1, Cases 7-12 on Slice 2, Cases 13-18 on Slice 3, Cases 19-24 on Slice 4, and Cases 25-30 on Slice 5, corresponding to the symbols on Fig. 2.

As mentioned earlier, the geometric constant, $K$, is an indicator of swirl strength, with a higher value implying stronger azimuthal momentum. When $K$ becomes small, the swirling flow does not contain enough azimuthal momentum to produce a large spreading angle of the liquid film. This suggests that $K$ plays a pivotal role, with a slight change causing significant differences in flow dynamics. Therefore, the training process must preserve important flow physics during data reduction.

## Prediction by KSPOD-based emulation

Figure 4 compares the first three pressure POD modes between Cases 9 and 21 in Cluster B. Mode 1 is an ensemble of the longitudinal modes of the hydrodynamic instability. In both cases, the injector length is fixed at 25 mm, and the longitudinal hydrodynamic wave speed in the liquid film is estimated to be 7-9 m/s [8-10, 28]. This leads to a characteristic frequency of 0.32-0.45 kHz for the hydrodynamic instability. Mode 2 contains about 12% of the total energy and has similar structures similar Cases 9 and 21, except in the downstream region close to the centerline. Mode 3 has around 5% of the total energy, and the distributions for the two cases are alike. The dominant frequencies for mode 2 in the two cases are 0.61 and 0.63 kHz, respectively. For mode 3, dominant frequencies of 1.14 and 1.39 kHz are observed. The frequencies associated with the recirculating flow downstream of the injector exit are 0.65 and 1.29 kHz, respectively. Hence, modes 2 and 3 can be attributed to excitation by the precessing vortex core. In short, modes 1-3 capture similar physical characteristics in the dominant POD modes with the same order resulting from the eigen-decomposition. The similarity of POD modes among cases in the same cluster justifies the two fundamental assumptions of KSPOD stated on Page 3.

Figure 5 shows the accumulated energy percentage of the density POD modes for the cases presented in Fig. 3. The first 150 modes contain over 90% of the total energy, and the first 300 modes have over 99.9% of the energy. KSPOD builds a posterior model based on the basis functions ranked by the eigen-decomposition. As such, the reduced data sets (that is, the dominant modes) capture the significant flow structures and their dynamic characteristics for the same rank.

The 30 cases selected by DoE have different inlet velocities, which generate different instability wave speeds. Phase information, however, cannot be reproduced perfectly, due to the presence of turbulence. If two designs with similar dynamic mechanisms, such as Cases 4 and 6, have phase differences of 90° or 180°, the prediction results may be excessively smoothed by traditional kriging. If the weighting numbers evaluated by kriging for these two cases are close, and their instability waves are exactly out of phase, the wave information cancels out. As such, KSPOD applies a weighting function for the POD modes to ensure that similar POD modes in different cases can retain appropriate phase information.

Figure 6 shows excellent comparison between the simulation result and prediction from the trained emulator for Test Case A1. The design parameters are given in Table 5. The KSPOD-based emulator works well; it is able to capture essential flow structures successfully. The evolution of the liquid film and its spreading downstream of the injector exit also agree well between the simulation and emulation. The overall turnaround time for emulation prediction, excluding data loading and training, is about 42 seconds of CPU time for one snapshot. Table 6 lists the contribution of POD modes from each case. The weighting number is calculated based on the Test Case A1. It is noted that Cases 17 and 30 provide respectively 91.77% and 13.37% weighting, while Cases 5, 10, 22, and 28 provide over 10% negative weights on A1.

Figure 7 shows prediction results for the four different design clusters, in each of which two test cases are considered. The flow structures and dynamics are well captured by the emulator. In Case C1, the corner recirculation near the headend of the injector is clearly observed. The discrepancy caused by time delay is present in Cases A1 and B2. In A1, the traveling surface wave in the injector propagates downstream slightly faster in the simulation than in the emulation.

Table 5. Design parameters for eight test cases in four different clusters

| Case | $\delta$ (mm) | $\theta$ (deg) | $\Delta L$(mm) | $u_{in}$ (m/s) | $u_r$ (m/s) | $u_\theta$ (m/s) | Cluster |
|---|---|---|---|---|---|---|---|
| A1 | 1.26 | 44.11 | 0.94 | 6.55 | 4.70 | 4.56 | A |
| A2 | 1.20 | 41.97 | 0.90 | 6.65 | 4.94 | 4.44 | A |
| B1 | 0.70 | 40.73 | 2.71 | 11.12 | 8.43 | 7.26 | B |
| B2 | 0.71 | 52.59 | 3.24 | 13.79 | 8.38 | 10.95 | B |
| C1 | 0.42 | 37.73 | 2.41 | 17.91 | 14.16 | 10.96 | C |
| C2 | 0.49 | 57.12 | 2.88 | 22.33 | 12.12 | 18.75 | C |
| D1 | 0.27 | 50.39 | 1.40 | 34.37 | 21.91 | 26.48 | D |
| D2 | 0.33 | 60.76 | 2.32 | 36.32 | 17.74 | 31.70 | D |

Table 6. Weighting numbers from of POD modes for Case A1

| Case | 1 | 2 | 3 | 4 | 5 | 6 | 7 | 8 | 9 | 10 |
|---|---|---|---|---|---|---|---|---|---|---|
| Cluster | D | B | B | A | A | A | C | C | B | A |
| Weighting Number | -0.01% | 0.50% | 2.13% | 0.82% | -4.24% | 0.02% | 0.00% | 0.12% | 0.04% | -2.18% |
| Case | 11 | 12 | 13 | 14 | 15 | 16 | 17 | 18 | 19 | 20 |
| Cluster | A | A | C | B | B | A | A | A | D | B |
| Weighting Number | 2.17% | -0.03% | 0.16% | 0.18% | -0.13% | 0.00% | 91.77% | -0.17% | -0.04% | -0.08% |
| Case | 21 | 22 | 23 | 24 | 25 | 26 | 27 | 28 | 29 | 30 |
| Cluster | B | A | A | A | C | C | B | A | A | A |
| Weighting Number | 0.16% | -2.50% | 0.20% | -0.13% | 0.05% | 0.01% | 0.00% | -2.28% | 0.08% | 13.37% |

## Accuracy of prediction: film thickness and spreading angle

To further evaluate the accuracy of the KSPOD-based emulation, two performance measures, film thickness and spreading angle, are assessed. Table 7 presents a comparison of the time-mean simulation and emulation (prediction) results, obtained by averaging the instantaneous data over a statistically meaningful time duration. The error is calculated as follows.

$$\varepsilon_{abs} = \frac{|x_{sim} - x_{emu}|}{x_{sim}} \times 100\% \quad (7)$$

where $x_{sim}$ represents data from simulation and $x_{emu}$ from emulation. As most training cases are located in Clusters A and B, the predictions are more accurate within these two clusters.

Table 7. Film thickness and spreading angle for simulation and emulation results

| | Spreading Angle | | | Film Thickness | | |
|---|---|---|---|---|---|---|
| | Simulation | Emulation | $\overline{\varepsilon_{abs}}$ | Simulation | Emulation | $\overline{\varepsilon_{abs}}$ |
| Case A1 | 52.85 | 52.92 | 0.14% | 0.629 | 0.625 | 0.51% |
| Case A2 | 52.57 | 51.96 | 1.15% | 0.637 | 0.657 | 3.14% |
| Case B1 | 54.22 | 53.66 | 1.02% | 0.582 | 0.600 | 3.03% |
| Case B2 | 53.81 | 53.87 | 0.12% | 0.594 | 0.592 | 0.40% |
| Case C1 | 57.68 | 57.71 | 0.05% | 0.469 | 0.473 | 0.85% |
| Case C2 | 57.78 | 57.74 | 0.06% | 0.475 | 0.472 | 0.63% |
| Case D1 | 59.00 | 58.03 | 1.64% | 0.379 | 0.378 | 0.26% |
| Case D2 | 61.59 | 61.33 | 0.41% | 0.370 | 0.377 | 1.97% |

The probability densities of instantaneous spreading angle and liquid film thickness for four selected training cases in Cluster A are obtained from the estimated kernel smoothing function, as shown in Fig. 8. A kernel distribution is a nonparametric representation of the probability density function, $f_h(x)$, of a random variable, written as

$$f_h(x) = \frac{1}{nh} \sum_{i=1}^{n} \mathcal{K}\left(\frac{x - x_i}{h}\right) \quad (8)$$

where $n$ is the sample size, $\mathcal{K}(\cdot)$ the density smoothing function, and $h$ a smoothing parameter named bandwidth. The distributions of probability density for cases in the same cluster bear close similarities, but do not collapse.

Figure 9 compares the probability density distributions of spreading angle and liquid film thickness between simulations and emulations for all test cases. The vertical lines represent mean values. Detailed information about mean values, standard deviations, and averaged absolute error $\overline{\varepsilon_{abs}}$ is given in Table 7. The maximum absolute errors for liquid film thickness and spreading angle are 1.64% in Case D1 and 3.14% in Case A2, respectively. The liquid film thicknesses are relatively small in Clusters C and D, so their mean values visually overlap each other.

Another way to measure the performance of the emulation is based on the distribution of liquid film thickness along the axial direction. Figure 10 shows a comparison between the simulation and emulation results, averaged over 1000 snapshots. Figure 11 shows the absolute error between simulations and emulations. The horizontal lines represent averaged absolute error $\overline{\varepsilon_{abs}}$ for each test case. Overall, the liquid film thickness predicted by the KSPOD-based emulation has an averaged error less than 5%, except for Case B2, which has $\overline{\varepsilon_{abs}} = 6.4\%$. Cases with higher inlet velocities have less variation for the liquid film thickness near the injector exit. The first local maximum of error occurs in the LOX inlet area, the region that contains the highest momentum and kinetic energy. The second peak of error takes place when the flowfield is still developing. The large error along the axial direction in Case B2 can be attributed to the fact that this case is close to the boundary between Clusters B and C, and includes more flow mechanisms that are prominent in the cases in Cluster C. The 30 cases used for model training are mostly distributed in Clusters A and B. If the 30 training cases had space-filling properties with respect to inlet velocities, the prediction error could be decreased and the quantitative analysis for Test Cases B2, C1, and C2 could be improved.

## IV. Conclusion

This study develops statistical methods for building an efficient and accurate emulation (surrogate) model for the prediction of spatiotemporal flow dynamics. The emulation method is constructed on a kernel-smoothed proper-orthogonal-decomposition (KSPOD) technique, which leverages kriging-based weighted functions from the design matrix. As an example, the spatiotemporal flow evolution in a swirl injector is investigated over a wide range of design parameters and operating conditions. The KSPOD-based emulation model is validated against high-fidelity simulation results obtained from large-eddy simulations (LES). The model not only preserves key physical mechanisms underlying the flow evolution, but quantitatively captures the dynamics over a wide range of temporal and spatial scales of concern. In addition, the model enables effective design surveys utilizing high-fidelity simulation data, achieving a turnaround time for evaluating new design points that is 42,000 times faster than the original simulation.

Based on the current results, further investigation is necessary to improve the emulation model, data reduction, and feature extraction of flow dynamics. The KSPOD could also be combined with common-grid POD (CPOD) techniques to mitigate the uncertainties of prediction for a broad-range design space. Moreover, the use of artificial neural networks could potentially be employed, if an activation function that properly treats the physics of the problem can be identified. More effective incorporation of physical knowledge for model tuning should also be explored and implemented.

## Acknowledgement

This work was sponsored partly by the William R. T. Oakes Endowment and partly by the Coca-Cola Endowment of the Georgia Institute of Technology.

# Appendix

## Interpretation as a Nadaraya-Watson kernel smoother

Suppose simulations are conducted at various design geometries $\boldsymbol{c} = \{\boldsymbol{c_1}, \dots, \boldsymbol{c_n}\}$, and it is assumed that the true function $W(\boldsymbol{c})$ is a realization from a stochastic process

$$W(\boldsymbol{c}) = \mu + Z(\boldsymbol{c}) \tag{9}$$

where $\boldsymbol{c}$ is an $n$-dimentional vector (with $d$ design variables), $\mu$ a constant global model, and $Z(\boldsymbol{c})$ a local deviation from the global model with zero mean. Consider the kriging of the indicator vector $\boldsymbol{e}_i$, an $n$-vector with unity in the $i$-th entry, and zero elsewhere. There is another vector $\boldsymbol{r_i}$, where $i = 1, \cdots, p$ and $p$ is the number of control settings for design cases. Since a space-filling design is employed, it is easy to show that the optimal correlation parameters for the underlying Gaussian process should be equal for all $p$ dimensions. Denote this common correlation as $\theta$. When the number of design points $n \to \infty$, one can show that $\theta \to \infty$ as well, since the "kriging surface" for $\boldsymbol{e}_i$ converges pointwise to a discontinuous surface with value 1 at $\boldsymbol{r}_i$ and 0 elsewhere. The kriging estimate $\boldsymbol{e}_i$ for a new design setting $\boldsymbol{r}_{new}$ is

$$\widehat{w}_{new,i} = \hat{\mu} + \boldsymbol{r}_{new}^T \boldsymbol{R}^{-1}(\boldsymbol{e}_i - \hat{\mu}\mathbf{1}_n) \tag{10}$$

where $\hat{\mu} = (\mathbf{1}_n^T \boldsymbol{R}^{-1} \mathbf{1}_n)^{-1} \mathbf{1}_n^T \boldsymbol{R}^{-1}$, $\boldsymbol{e}_i = \boldsymbol{R}_{ii}^{-1} / \sum_{i=1}^{n} \boldsymbol{R}_{ii}^{-1}$, $\boldsymbol{R} \equiv Corr\left[Z(\boldsymbol{c}_i), Z(\boldsymbol{c}_j)\right]$ are $n \times n$ correlation matrices, and $\widehat{w}_{new,i}$ a weighted number based on kriging. When $\theta \to \infty$ and $n \to \infty$, the best linear unbiased predictor estimator $\hat{\mu} \to 0$; when $\theta \to \infty$, the inverse correlation matrix $\boldsymbol{R}^{-1}$ converges element-wise to $\mathbf{I}_n$. Under these two approximations, we get a new kernel

$$k_\theta(\boldsymbol{c}_i, \boldsymbol{c}_{new}) = \widehat{w}_{new,i} \approx \boldsymbol{r}_{new}^T \boldsymbol{R}^{-1} \boldsymbol{e}_i = \exp\left\{-\theta \left|\left|\boldsymbol{c}_i - \boldsymbol{c}_{new}\right|\right|_2^2\right\} \tag{11}$$

where $||\cdot||_2$ is the Euclidean norm. In other words, $\widehat{w}_{new,i}$ is the isotropic Gaussian kernel $k_\theta(\boldsymbol{c}_i, \boldsymbol{c}_{new})$. The proposed predictor of the first $k$-th mode at the new design setting $\boldsymbol{c}_{new}$ is

$$\hat{\phi}_{new}^k(\boldsymbol{x}) = \frac{\sum_{i=1}^n \widehat{w}_{new,i} \phi_i^k(\boldsymbol{x})}{\sum_{i=1}^n \widehat{w}_{new,i}} \approx \frac{\sum_{i=1}^n k_\theta(\boldsymbol{c}_i, \boldsymbol{c}_{new}) \phi_i^k(\boldsymbol{x})}{\sum_{i=1}^n k_\theta(\boldsymbol{c}_i, \boldsymbol{c}_{new})} \tag{12}$$

where $\phi_i^k(\boldsymbol{x}), i = 1, \cdots, n$ – The $k$-th POD mode at design setting $\boldsymbol{c}_i$.

With kriging over $k$ modes, the new POD coefficient is defined as

$$\hat{\beta}_{new}^k(\boldsymbol{x}, t) = \left(\hat{\beta}_1^k, \hat{\beta}_2^k, \dots, \hat{\beta}_{10d}^k\right) \tag{13}$$

and the corresponding POD expansion using $k_{th}$ modes (that is, the prediction) is given by

$$\hat{f}_{new}(\boldsymbol{x}, t) = \sum_{k=1}^{\infty} \hat{\beta}_{new}^k(\boldsymbol{x}, t) \hat{\phi}_{new}^k(\boldsymbol{x}) = \sum_{k=1}^{\infty} \frac{\beta_i^k(x,t)\left(\sum_{i=1}^n k_\theta(c_i, c_{new}) \phi_i^k(x)\right)}{\left(\sum_{i=1}^n k_\theta(c_i, c_{new})\right)^2} \tag{14}$$

Eq. (11) can be seen as a kernel smoother on the observed modes $\left\{\phi_i^k(\boldsymbol{x})\right\}_{i=1}^n$ and coefficients $\left\{\beta_i^k(\boldsymbol{x}, t)\right\}_{i=1}^n$.

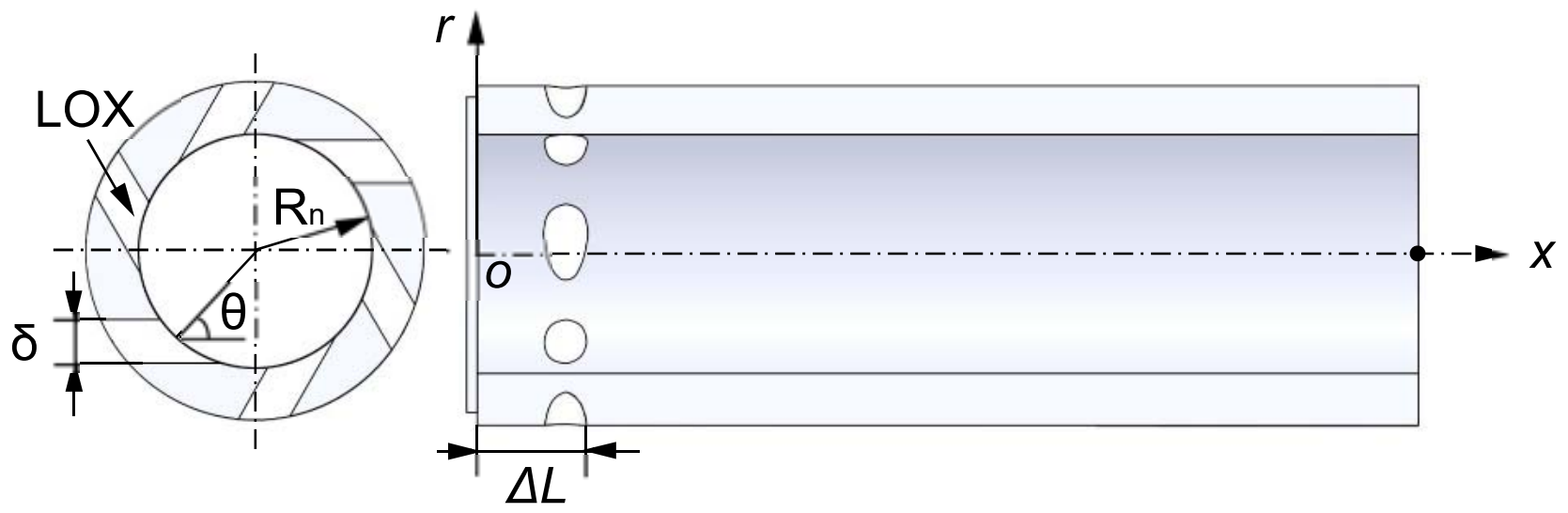

Figure 1. Schematic of swirl injector

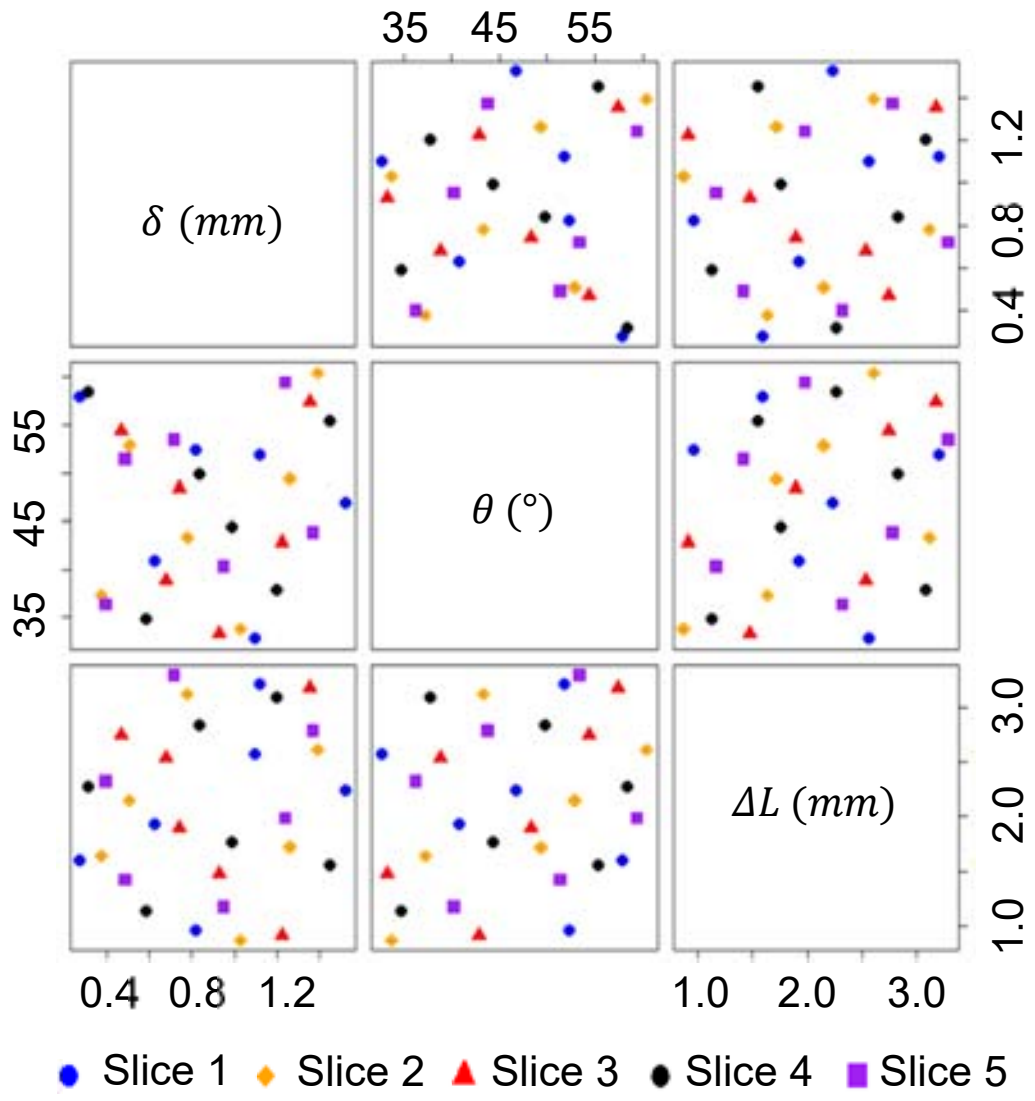

Figure 2. Two-dimensional projection of design points obtained by Sliced Latin Hypercube Design (SLHDs) methodology in the design space.

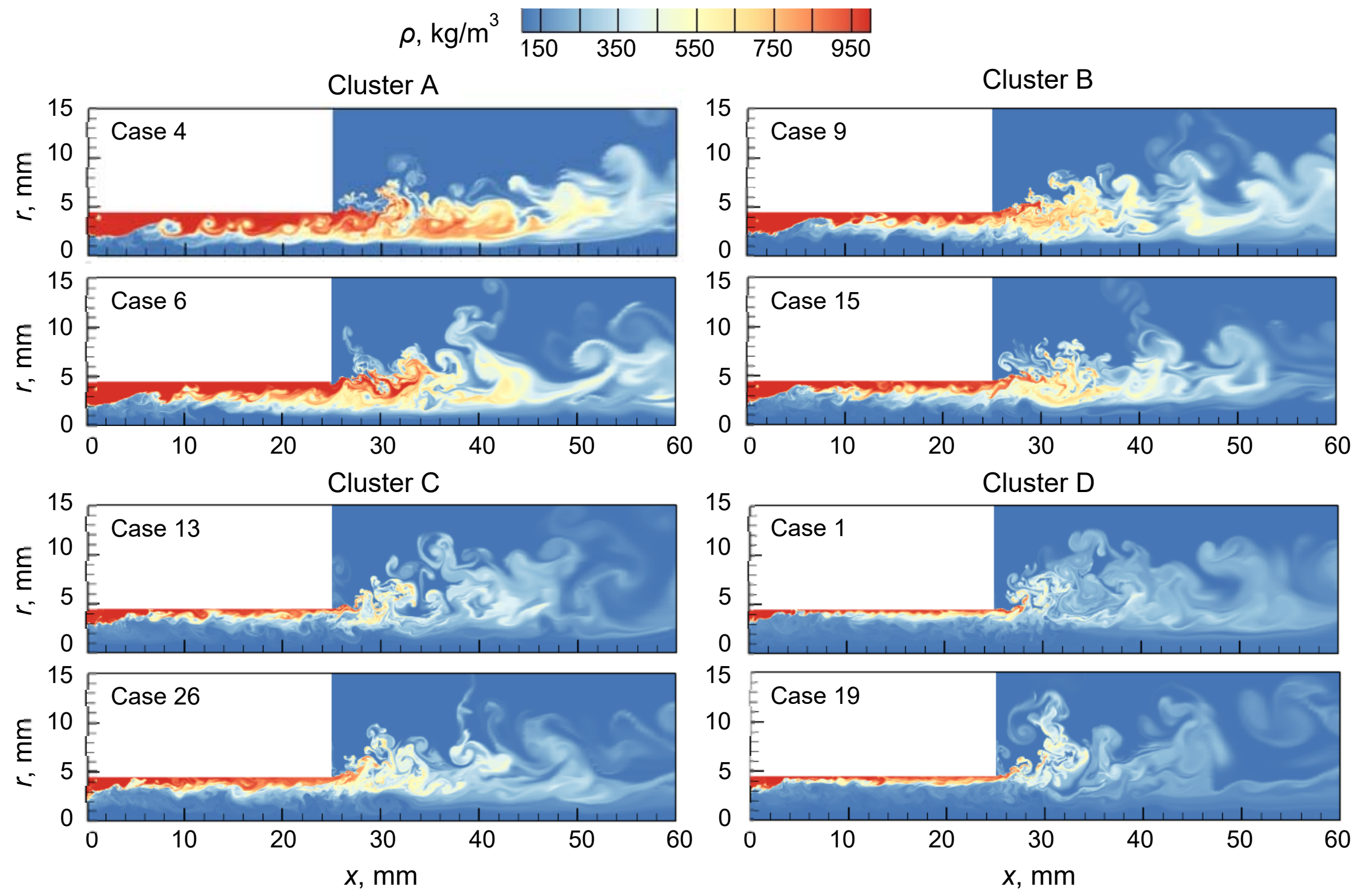

Figure 3. Snapshots of density field in different clusters at $t$ = 1.21 *ms*, obtained from LES-based high fidelity simulations.

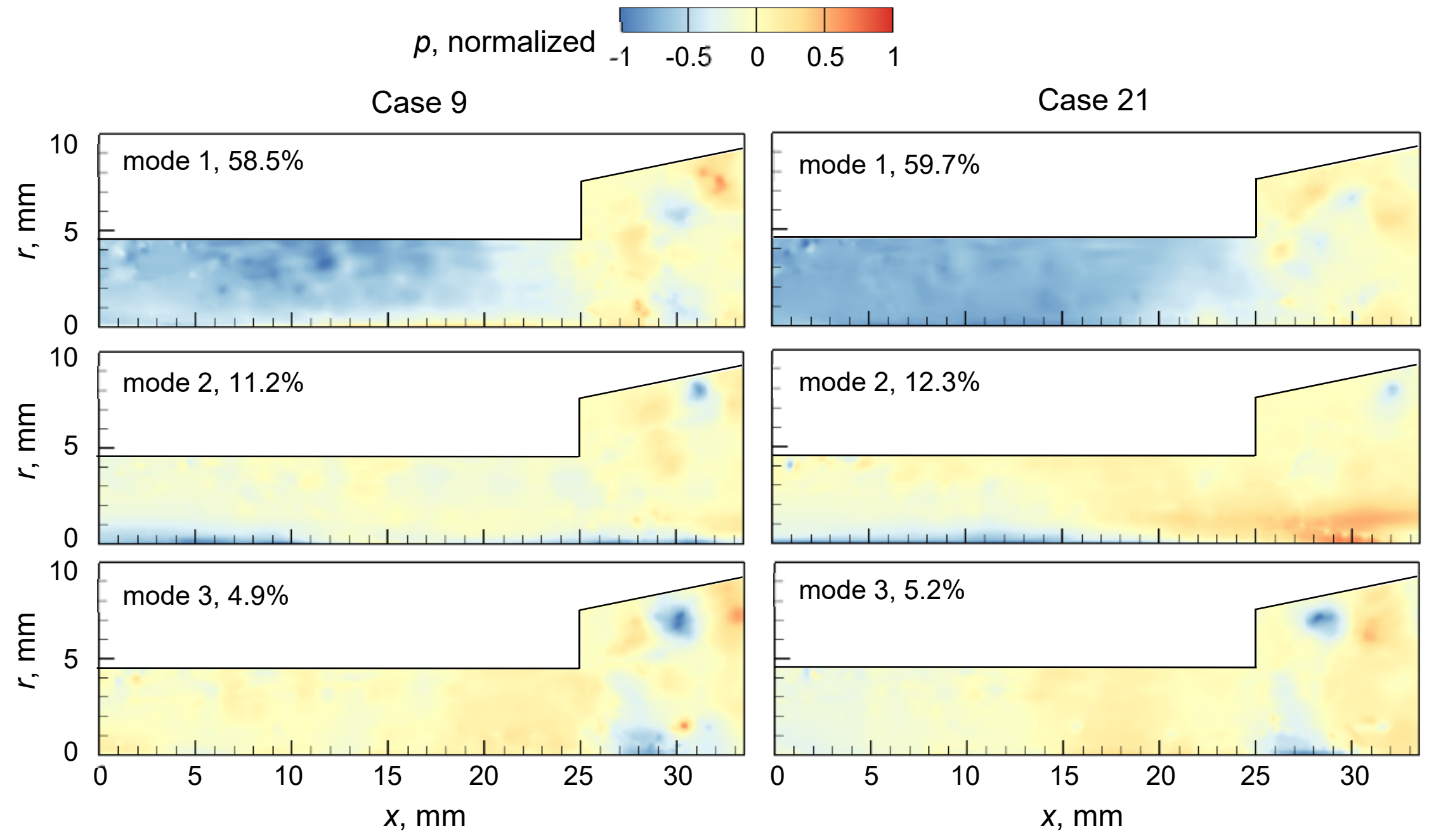

Figure 4. Pressure POD modes 1, 2, and 3 for Cases 9 and 21 from Cluster B.

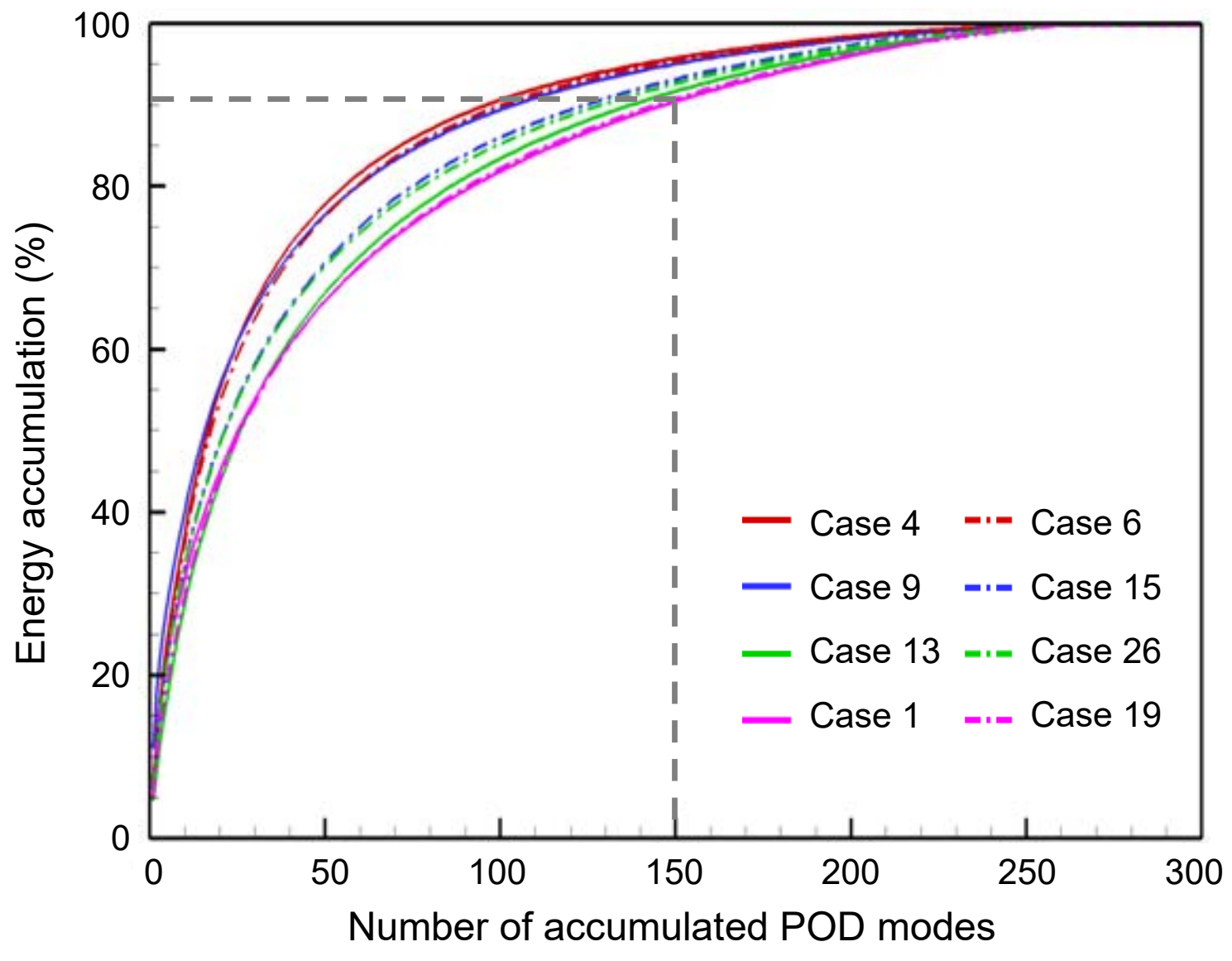

Figure 5. Energy accumulation of density POD modes

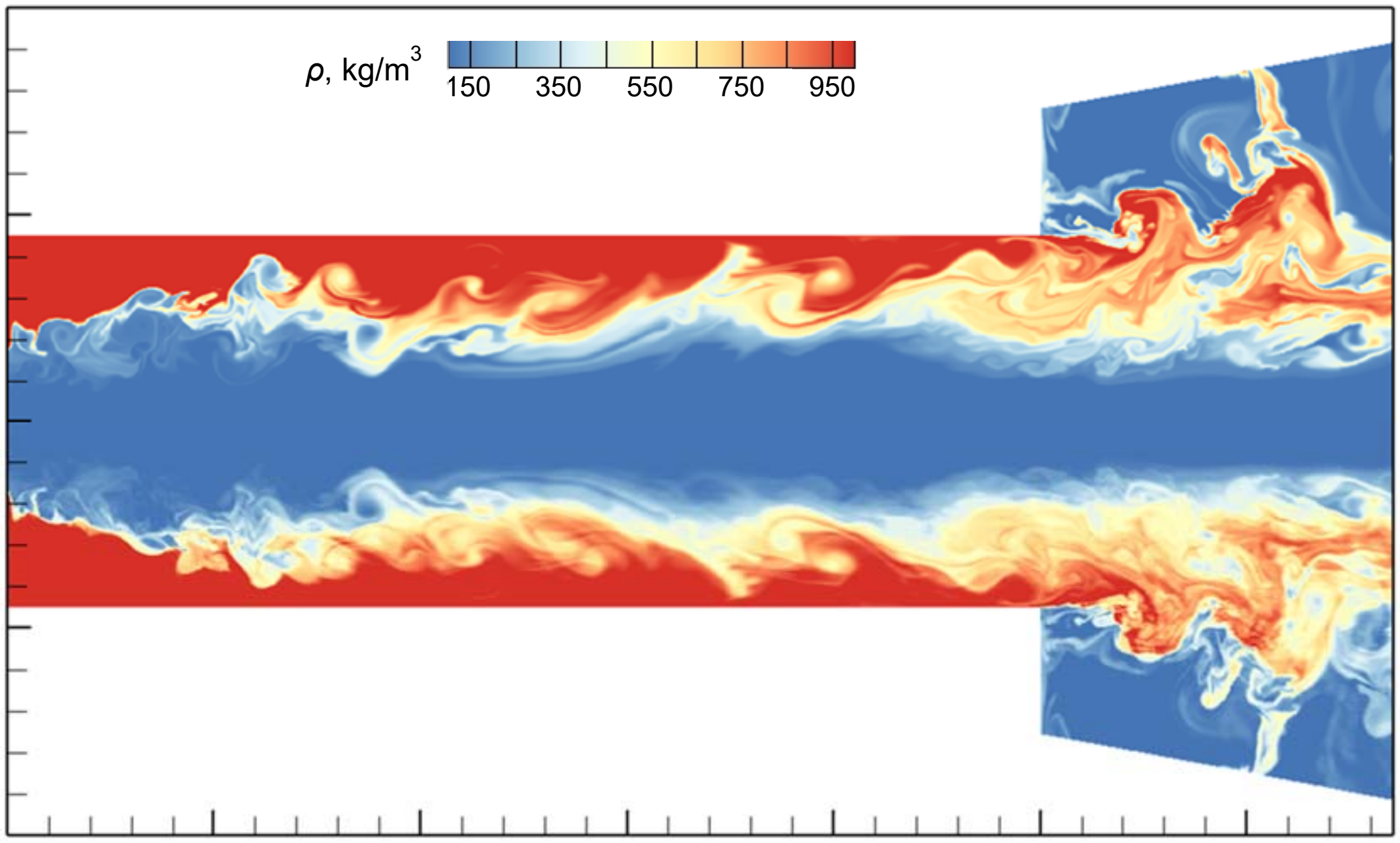

Figure 6 Comparison of density field between LES-based simulation and prediction by KSPOD-based emulation. Test Case A1.

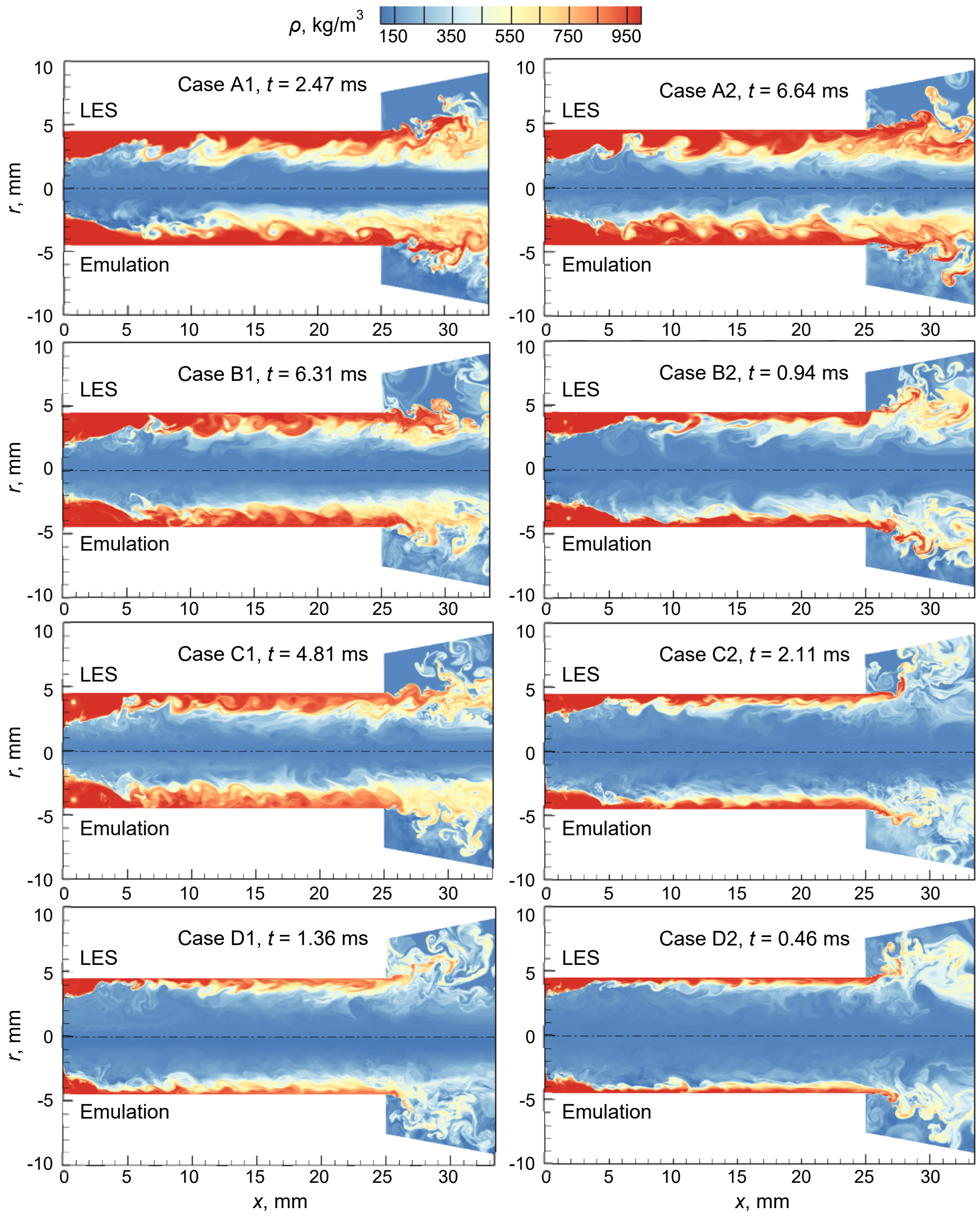

Figure 7. Comparison of density field between LES-based simulation and prediction by KSPOD-based emulation.

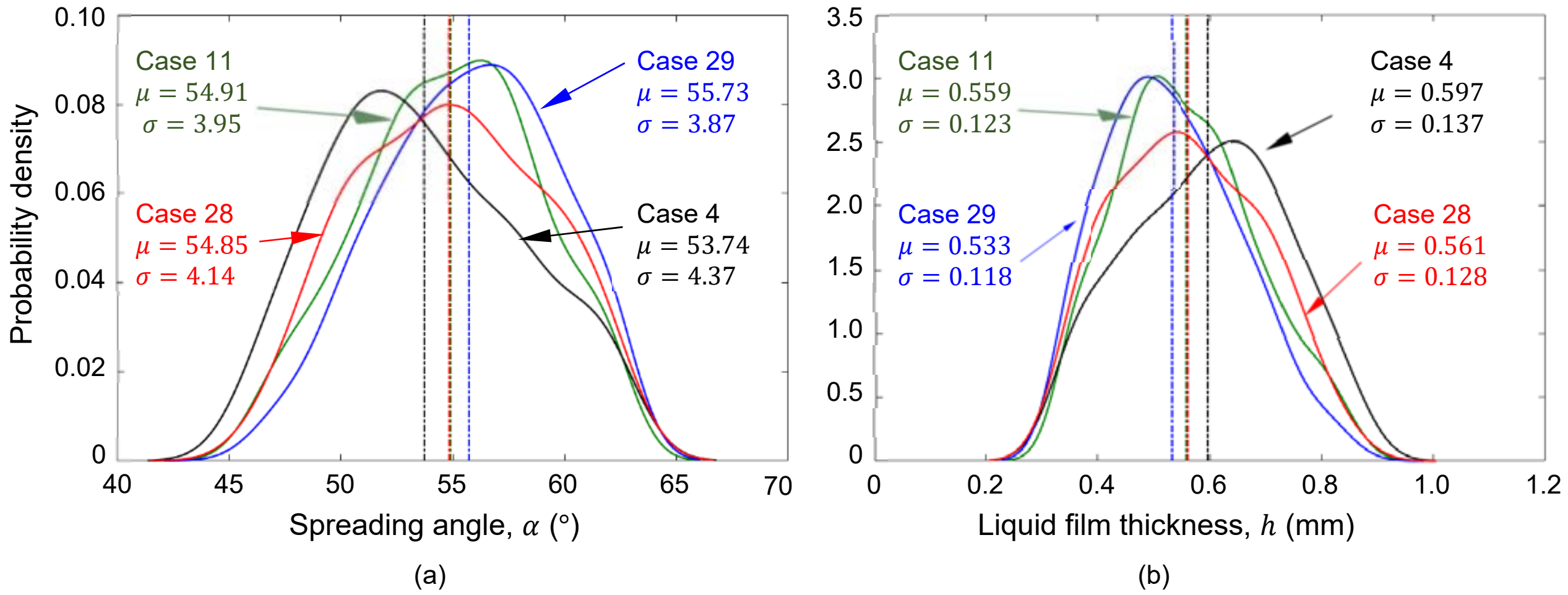

Figure 8. Probability densities of instantaneous spreading angle and liquid film thickness for Cases 4, 11, 28 and 29 from Cluster A. Vertical lines represent mean values.

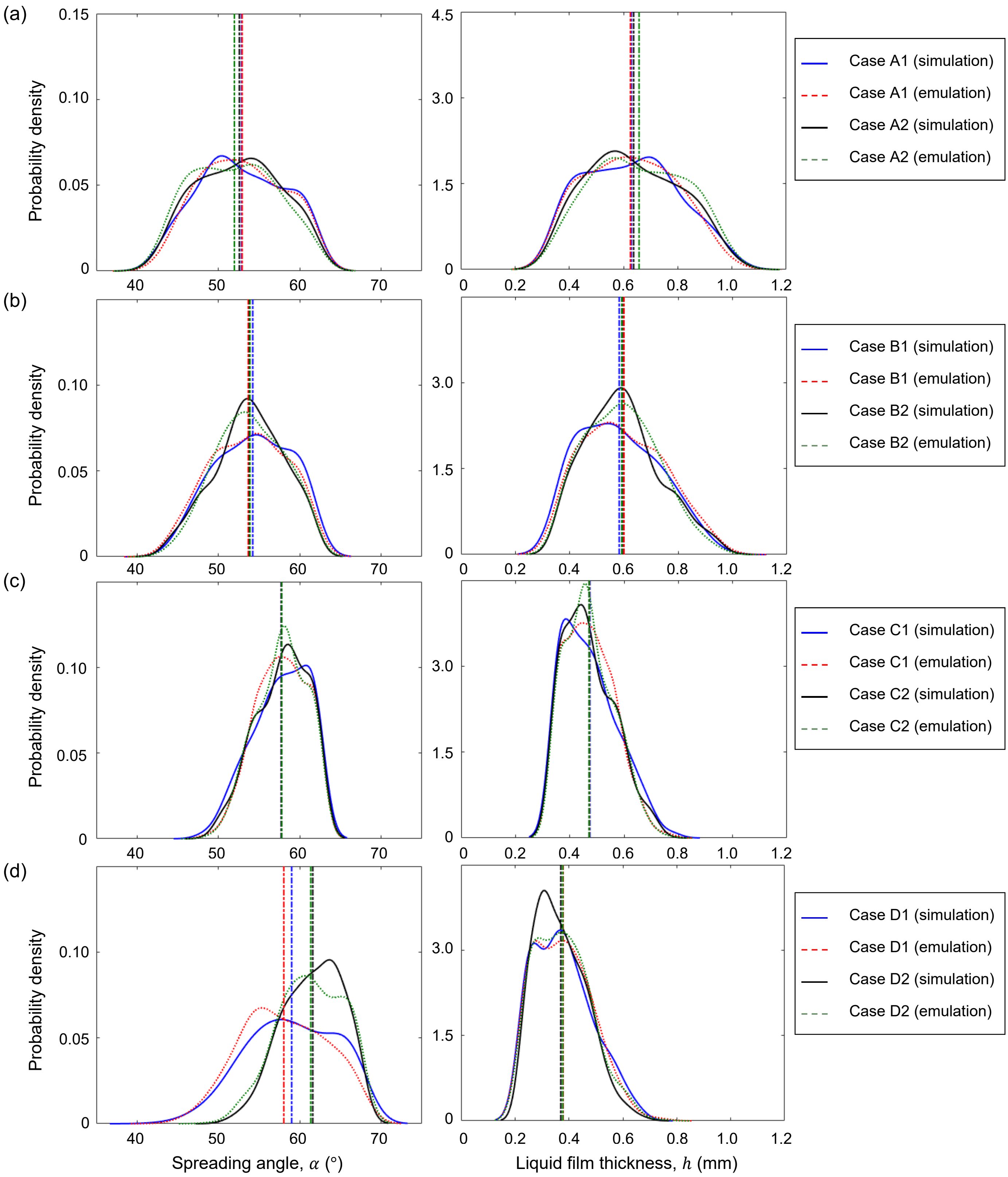

Figure 9. Probability densities of instantaneous spreading angle and liquid film thickness for test cases. Vertical lines represent mean values.

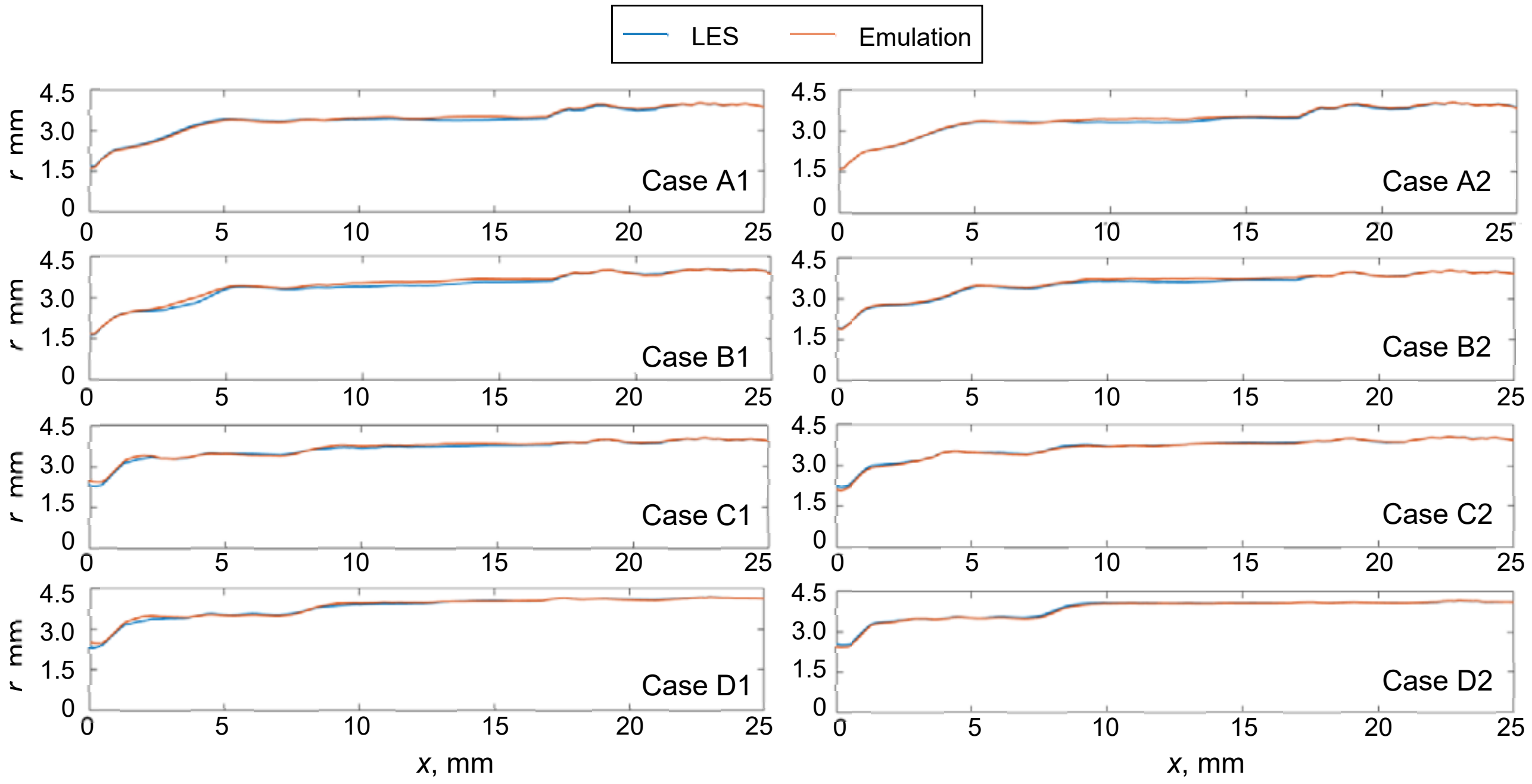

Figure 10. Comparison of liquid film thickness along the axial direction, averaged over 1000 snapshots.

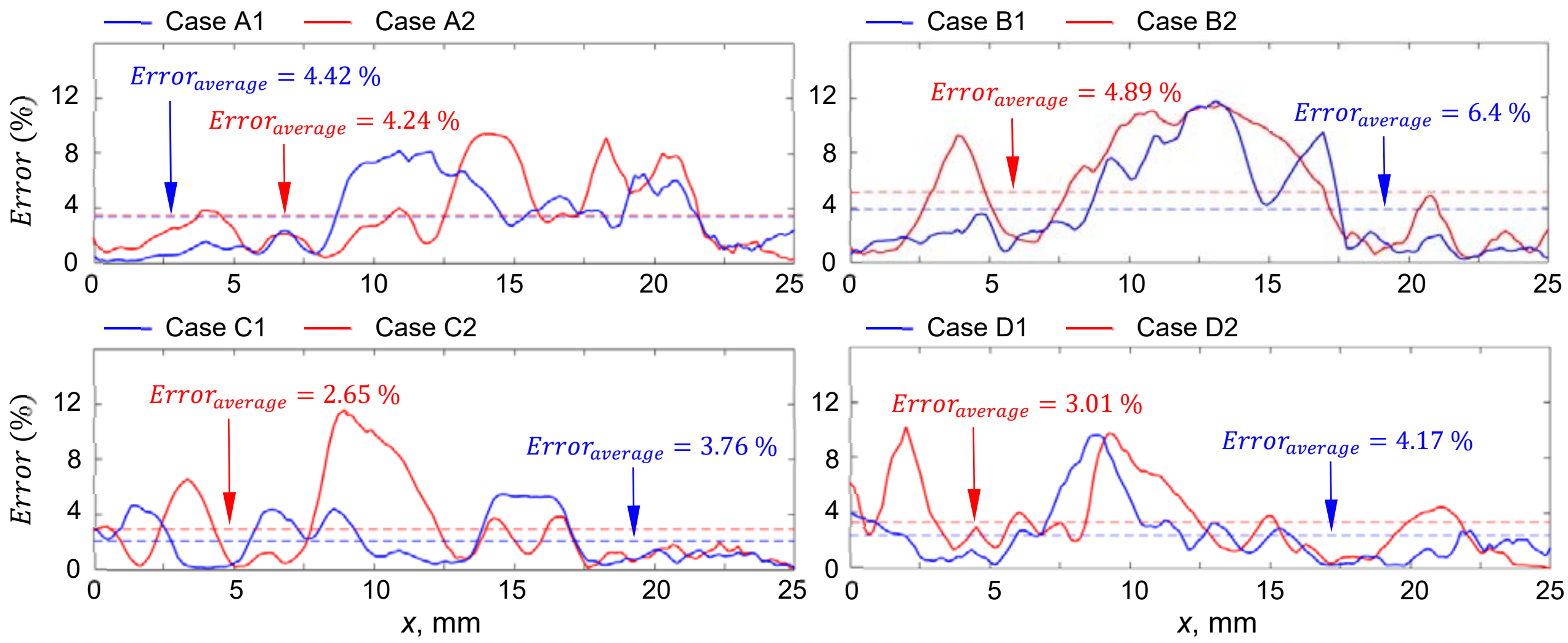

Figure 11. Error for liquid film thickness along the axial direction.